\documentclass[12pt]{article}

\usepackage{amsmath}
\usepackage{amssymb}
\usepackage{amsfonts}
\usepackage{amsthm}
\usepackage{color}
\usepackage{latexsym}
\usepackage{color,soul} 
\usepackage{xcolor}
\usepackage{ulem}
\usepackage{braket}
\usepackage{indentfirst}
\usepackage{cite}

\DeclareSymbolFont{matha}{OML}{txmi}{m}{it}
\DeclareMathSymbol{\varv}{\mathord}{matha}{118}

\def\det{\,\mbox{det}\,}

\def\al{\alpha}
\def\be{\beta}
\def\ga{\gamma}
\def\Ga{\Gamma}
\def\de{\delta}

\def\vp{\varepsilon}

\def\la{\lambda}

\def\si{\sigma}

\def\om{\omega}

\def\pa{\partial}

\def\beq{\begin{eqnarray}}
\def\eeq{\end{eqnarray}}

\definecolor{Purple}{rgb}{0.4,0.3,0.}

\begin{document}

\begin{center}

{\Large
Background field method and nonlinear gauges
}
\vskip 8mm

{\large
Breno L. Giacchini$^{ab}$, \ \
Peter M. Lavrov$^{cbd}$
\ \ and \ \
Ilya L. Shapiro$^{bcd}$}

\end{center}
\vskip 1mm

\begin{center}
{\sl
(a) \ 
Department of Physics, Southern University of Science and Technology,
\\
Shenzhen 518055, China
\vskip 3mm

(b) \ Departamento de F\'{\i}sica, ICE,
Universidade Federal de Juiz de Fora
\\
Juiz de Fora, CEP: 36036-330, MG,  Brazil
\vskip 3mm

(c) \  Department of Mathematical Analysis, Tomsk State
Pedagogical University,
\\ 
634061, Kievskaya St. 60, Tomsk, Russia
\vskip 3mm

(d) \ National Research Tomsk State  University,
\\
Lenin Av.\ 36, 634050 Tomsk, Russia
}
\vskip 2mm\vskip 2mm

{\sl E-mails:
\ \
breno@sustech.edu.cn,
 \
 lavrov@tspu.edu.ru,
\
shapiro@fisica.ufjf.br }

\end{center}
\vskip 6mm

\begin{abstract}
\noindent
We present a reformulation of the background field method for
Yang-Mills type theories, based on using a superalgebra of generators
of BRST and background field transformations. The new approach
enables one to implement and consistently use non-linear gauges in
a natural way, by using the requirement of invariance of the fermion
gauge-fixing functional under the background field transformations.
\end{abstract}

\begin{quotation}
\end{quotation}


\section{Introduction}
\label{Sec.1}

The standard approach to the Lagrangian quantization of gauge
theories \cite{FP,BV1,BV2} assumes the violation of the gauge invariance
of the action. In some cases, \textit{e.g.}, in the semiclassical gravity
theory \cite{RenCurved}, this makes all considerations and also
practical calculations quite complicated. The same certainly
concerns quantum gravity. Fortunately, there is a useful approach
to the quantization of gauge theories, known as the background field
method (BFM) \cite{DeW,AFS,Abbott}. Within the BFM one can explicitly
preserve the gauge invariance of an effective action in
all stages, and thus all physical results are reproduced using the
effective action of background fields (background effective action).
In many cases, the calculations in gauge theories in flat and curved
space-time are performed by means of the BFM,
while the general theorems concerning gauge invariant
renormalizability and gauge fixing dependence rely on the
standard version of quantization. This situation makes it at least
highly desirable to construct a completely consistent formulation
of the BFM for gauge theories and quantum gravity, such that
the mentioned general theorems could be formulated directly
within the formalism that is used for calculations and, \textit{e.g.}, for
the analysis of power counting in new non-conventional
models of quantum field theory (see, for instance, \cite{CountGhost,LSh}).

The BFM has been the object of extensive investigation from different
viewpoints (see \textit{e.g.}
\cite{K-SZ,GvanNW,CMacL,IO,GS,Ven,Reuter,Gr,BC,FPQ}), and
in the last years the interest in this method has grown rapidly,
since it is supposed to solve many important problems of gauge
theories
\cite{BQ,Barvinsky:2017zlx,BLT-YM,FT,Lav,BFMc,BLT-YM2,LSh}.
One of the standard assumptions of the BFM is related to the special
choice of gauge generators and gauge-fixing conditions, that are
typically linear in the quantum fields. Indeed, even though the gauge fixing
may admit a lot of arbitrariness, linearity is always assumed
in the framework of the BFM. At the same time, the consistent
formulation of the quantization method should certainly be free of
such a restriction, as the choice of the gauge fixing is arbitrary
and is not related to the physical contents of the initial gauge
theory. In the present paper we propose a generalization of the
standard formulation of the BFM, in the sense that is can be
used for a wide class of non-linear gauge-fixing conditions
and together with some types of non-linear gauge generators.
This new point of view on the BFM is based on the superalgebra of
generators of the fundamental symmetries of this formalism (namely, the
BRST \cite{BRS1,T} and the background field symmetries
\cite{DeW,AFS,Abbott}).

It is well known that a judicious choice of gauge condition can
yield a considerable simplification of calculations in quantum
field theory, and non-linear gauges are often used with this
purpose (see, for instance, \cite{Gervais:1972tr,Fujikawa:1973qs,Das:1980zy,Gavela:1981ri,
Ven,Raval:2016nsz}).
On the other hand, such gauges
emerge naturally in the framework of the effective field theory
approach, after the massive degrees of freedom are integrated
out and we are left with the gauge theory of relatively light
quantum fields \cite{Weinberg80}. Thus, the consistent
implementation of non-linear gauge fixing in all approaches
to quantization looks relevant from many viewpoints.

The general analysis of renormalization of gauge theories under
non-linear gauge conditions has been presented in many papers, including
\cite{Hsu:1973yf,Shizuya:1976rz,
Tyutin:1981ws,Girardi:1982by,VLT82,ZinnJustin:1984dt}.
Nonetheless, non-linear gauges are not frequently used in the
background field formalism. A remarkable example of simultaneous use
of the BFM and a non-linear gauge
is the two-loop calculation in quantum gravity \cite{Ven},
which confirmed the previous calculation of~\cite{GS85}.
This correspondence is certainly a positive
signal, making even more clear the need for a general discussion on
this subject, which is not present in the literature.

The idea that the gauge-fixing function should transform in a
covariant way under background field transformations (a
condition that is trivially satisfied for the usual linear gauge), as
a form of preserving the background field invariance of the
Faddeev-Popov action, is present in many discussions on Yang-Mills
theories~\cite{Grassi:2003mv,Barvinsky:2017zlx}. Nevertheless, the
applicability of the BFM with non-linear gauges is not a consensus.
For example, in~\cite{Bornsen:2002hh} it is mentioned that the
Gervais-Neveu gauge~\cite{Gervais:1972tr} could be used within the
BFM for Yang-Mills theory, while in~\cite{Barvinsky:2017zlx} the
linearity of the gauge-fixing condition and gauge generators is
regarded as an important factor for the consistency of the quantum
formalism.

As it was already mentioned above, in the present work we introduce
a new geometric point of view on the BFM based on the operator
superalgebra which underlies the method.
As the main application of this formalism, we obtain
necessary conditions for the consistent application of the BFM
\cite{DeW,AFS,Abbott} for Yang-Mills type theories with
non-linear gauge-fixing conditions and non-linear gauge generators.
The work is organised as follows. In Sec.~\ref{Sec.2} we briefly
review the BFM applied to gauge theories. The main original results
are presented in Sec.~\ref{Sec.3}, including the discussion of
non-linear gauges for generalised Yang-Mills theories in the BFM.
In Sec.~\ref{Sec.4} some aspects of the
Yang-Mills theory and quantum gravity are explicitly considered as examples of the
general result. Finally, in Sec.~\ref{Sec.5} we draw our conclusions.
The DeWitt's condensate notations \cite{DeWitt} are used
throughout the paper.

\section{Background field formalism for gauge theories}
\label{Sec.2}

As a starting point, consider a  gauge theory of fields $A^i$, with
Grassmann parity $\vp_i = \vp(A^i)$.  A complete set of
fields $\,A^i=(A^{\alpha k},A^m)\,$ includes fields $A^{\al k}$
of the gauge sector and also fields $A^m$ of the matter sector
of a given theory.

The classical action $S_0 (A)$
is invariant under gauge transformations
\beq
\delta_\xi A^i = R^i_\alpha(A) \xi^\alpha .
\eeq
Here $\xi^\alpha(x)$ is the transformation parameter with parity
$\vp(\xi^\alpha)=\vp_\alpha$, while $R^i_\alpha(A)$ (with
$\vp(R^i_\al) = \vp_i + \vp_\al$) is the generator of gauge
transformation. In our notation $i$ is the collection of internal,
Lorentz and continuous (spacetime coordinates) indices. We
assume that the generators are linearly independent, \textit{i.e.},
\beq
R^i_\alpha(A) z^\alpha = 0 \quad \Longrightarrow \quad z^\alpha = 0,
\eeq
and satisfy a closed algebra with structure coefficients
$F^\gamma_{\alpha\beta}$ that do not depend on the fields,
\beq \label{Comut}
[\delta_{\xi_1},\delta_{\xi_2}] A^i = \delta_{\xi_3} A^i
\qquad \text{with} \qquad
\xi^\al_3 = F^\al_{\be\ga}\, \xi^\ga_2 \xi^\beta_1 .
\eeq
From~\eqref{Comut} it follows that
$\,F^\ga_{\al\be}=-(-1)^{\vp_\al \vp_\be}F^\ga_{\be\al}$
and
\beq
\label{AlgebraAi}
R^i_{\al,j}(A)R^j_{\beta}(A) - (-1)^{\vp_\al \vp_\be}
R^i_{\be ,j}(A)R^j_{\al}(A) =  - R^i_\ga(A) F^\ga_{\al\be},
\eeq
where we denote the right functional derivative by $\de_r X / \de A^i = X_{,i}$.
In principle, the generators $R^i_\al(A)$ may be non-linear in the
fields. Further restrictions on the generators
will be introduced in
the next section, motivated by quantum aspects of the theory.

Let us formulate the theory within the BFM by splitting the original
fields $A^i$ into two types of fields, through the substitution $A^i
\longmapsto A^i + \mathcal{B}^i$ in the initial action $S_0(A)$. It
is assumed that the fields $\mathcal{B}^i$ are not equal to zero
{\it only} in the gauge sector
\footnote{
In gauge theories without
spontaneous symmetry breaking one can introduce background
fields in the sector of gauge fields without loss of generality
\cite{BLT-YM,BLT-YM2}.}.
These fields form a classical background,
while $A^i$ are quantum fields, that means being subject of
quantization, \textit{e.g.}, these fields are integration variables
in functional integrals. It is clear that the total action satisfies
\beq \label{VarYM-back} \de_\om S_0 (A+\mathcal{B}) = 0 \eeq under
the transformation $\,A^i \longmapsto A^{\prime i} = A^i + R^i_\al
(A+\mathcal{B}) \om^\al$. On the other hand, the new field
$\mathcal{B}^i$ introduces extra new degrees of freedom and, thence,
there is an ambiguity in the transformation rule for each of the
fields $A^i$ and $\mathcal{B}^i$. This ambiguity can be fixed in
different ways, and in the BFM it is done by choosing the
transformation laws \beq \label{BGTrans} \de_\om^{(q)} A^i = \left[
R^i_\al (A+\mathcal{B}) - R^i_\al (\mathcal{B}) \right] \om^\al ,
\qquad \de_\om^{(c)} \mathcal{B}^i = R^i_\al (\mathcal{B}) \om^\al,
\eeq defining the \textit{background field transformations} for the
fields $A^i$ and $\mathcal{B}^i$, respectively. The superscript
($q$) indicates the transformation of the quantum fields, while that
of the classical fields is labelled by ($c$). Thus, in
Eq.~\eqref{VarYM-back} one has $\de_\om = \de^{(q)}_\om +
\de^{(c)}_\om$. Indeed, the background field transformation rule for
the field $A^i$ was chosen so that \beq \de_\om^{(c)} \mathcal{B}^i
+ \de_\om^{(q)} A^i \,=\, R^i_\al (A+\mathcal{B}) \om^\al .
\label{backtrans} \eeq

In order to apply the Faddeev-Popov quantization scheme \cite{FP}
in the background field formalism,  one has to introduce a gauge-fixing condition
for the quantum fields $A^i$, and extra fields $C^\al$, $\bar{C}^\al$
and $B^\al$. To simplify notation we denote
by $\phi = \left\lbrace \phi^A \right\rbrace $ the set of all quantum fields
\beq
\phi^A =
\big( A^i, B^\al, C^\al, \bar{C}^\al \big).
\label{phi}
\eeq
The Grassmann parity of the fields
is $\,\vp(\phi^A) = \vp_A = (\vp_i,\vp_\al,\vp_\al+1,\vp_\al+1)$,
while their ghost numbers are
$\,\text{gh}(C^\al) = - \text{gh}(\bar{C}^\al)= 1\,$ and
$\,\text{gh}(A^i) = \text{gh}(B^\al)=0$.
 The corresponding Faddeev-Popov action
in the BFM reads
\beq
\label{S-FP_BFM}
S_{\text{FP}} (\phi,\mathcal{B})
\,=\, S_{0} (A+\mathcal{B}) + S_{\text{gh}} (\phi,\mathcal{B})
+ S_{\text{gf}} (\phi,\mathcal{B}),
\eeq
where the ghost and gauge-fixing actions are defined as
\beq
S_{\text{gh}} (\phi,\mathcal{B})
\,=\, \bar{C}^\al \, \frac{\de_r \chi_\al(A,\mathcal{B})}{\de A^i}
\, R^{i}_\be (A+\mathcal{B}) C^\be
\eeq
and
\beq
\label{Sgf-YM-BFM}
\quad
S_{\text{gf}} (\phi,\mathcal{B}) \,=\,
B^\al \Big( \chi_\al(A,\mathcal{B})
+ \frac{\xi}{2} g_{\al\be} B^\be \Big).
\eeq
In this expression $\,\xi\,$  is a gauge parameter that has to
be introduced in the case of a non-singular gauge condition, and
$g_{\al\be}$ is an arbitrary invertible constant matrix such that
$g_{\be\al}=g_{\al\be}(-1)^{\vp_\al \vp_\be}$.
We recall that in the BFM only the gauge of the quantum field is
fixed by $\chi_\al(A,\mathcal{B})$, while the symmetry for the
background fields may be preserved. The standard choice of
$\,\chi_\al(A,\mathcal{B})\,$ in the BFM is of the type
\beq
 \chi_\al(A,\mathcal{B}) \,=\,F_{\al i}(\mathcal{B})A^i,
\eeq
which is a gauge fixing condition linear in the quantum fields.

In the framework of Faddeev-Popov quantization, the gauge symmetry
of the initial action is replaced by the global supersymmetry (BRST
symmetry) of the Faddeev-Popov action (\ref{S-FP_BFM}), defined by
the transformation \cite{BRS1,T}
\beq
\de_{\text{BRST}}\,\phi^A
\,=\, \big( \hat{R}_{\text{BRST}} \, \phi^A \big) \la,
\eeq
where
$\la$ is a constant anticommuting parameter. The generator reads
\beq
\hat{R}_{\text{BRST}}(\phi,\mathcal{B}) \,=\, \frac{\de_r}{\de
\phi^A} \, R^A_{\text{BRST}}(\phi,\mathcal{B}), \label{BRSTgen}
\eeq
where
\beq
\label{R^i_B} R_{\text{BRST}}^A(\phi,\mathcal{B}) \,=\,
\Big( R^{i}_\al (A+\mathcal{B}) C^\al  , \,\, 0, \,\,  -\frac{1}{2}
F^{\al}_{\be\ga} C^\ga C^\be (-1)^{\vp_\be }, \,\,  (-1)^{\vp_\al}
B^\al  \Big) .
\eeq
Regarding the Grassmann parity, one has
$\vp(R^A_{\text{BRST}}(\phi,\mathcal{B}))=\vp_A+1$. The BRST
transformations are applied only on quantum fields, thus,
$\de_{\text{BRST}} \, \mathcal{B}^i = 0$.
Moreover, it is possible to show that the BRST
operator is nilpotent, \textit{i.e.},
\beq
\hat{R}_{\text{BRST}}^2(\phi,\mathcal{B}) = 0.
\eeq

Let us note that in the existing literature there is a formulation of
the BFM in terms of
extended BRST differentials \cite{Gr,BC,FPQ,BQ}, that coincides
with $\hat{R}_{\text{BRST}}(\phi,\mathcal{B})$ in the sector of
fields $\phi$. At the same time, the BRST variation of the background
fields is equal to  new ghost fields, so that in the extended field space
the usual BRST algebra takes place. In contrast to this approach, in
the next Section we propose a new superalgebra, underlying the BFM
which enables us to consider non-linear gauges and non-linear gauge
generators.

For the subsequent discussion, it is useful to introduce the fermion
gauge-fixing functional
\beq
\Psi(\phi,\mathcal{B})
\,=\,
\bar{C}^\al \chi_\al(A,B,\mathcal{B}),
\qquad
\mbox{with}
\qquad
\vp(\Psi)
\,=\,
- \text{gh} (\Psi) \,=\, 1.
\label{Psi}
\eeq
The Faddeev-Popov action can then be cast in the form
\beq
\label{FP-Psi}
S_{\text{FP}} (\phi,\mathcal{B})
\,=\, S_{0} (A+\mathcal{B})
+ \hat{R}_{\text{BRST}}(\phi,\mathcal{B})
\, \Psi(\phi,\mathcal{B}) .
\eeq
The BRST symmetry of $S_{\text{FP}}(\phi,\mathcal{B})$ can be easily
verified in this representation by applying the nilpotency property of the
operator $\hat{R}_{\text{BRST}}$.

Let us point out that in order to achieve economic notations, in
Eq.~(\ref{Psi}) and hereafter we let $\,\chi_\al\,$ to depend also on
the auxiliary field $B^\al$. This is a practically useful way of taking
into account the possibility of non-singular gauge conditions
(see Eq.~\eqref{Sgf-YM-BFM}). Nonetheless, further discussion
and consequent results do not require any kind of \textit{a priori}
specific dependence of the gauge-fixing functional on
$A^i, B^\al$ and $\mathcal{B}^i$.

\section{BFM compatible gauge functionals}
\label{Sec.3}

In this section we propose a new point of view and a generalization
of the standard BFM for gauge theories, that is based on using a
superalgebra of generators of all the symmetries underlying the method and works for
a wide class of gauge fixing conditions.

Apart from the global supersymmetry, a consistent formulation of the
BFM requires that the Faddeev-Popov action be invariant under
background field transformations. The former symmetry is ensured
in the representation~\eqref{FP-Psi} of the Faddeev-Popov action,
for any choice of gauge-fixing functional $\Psi$. Therefore, it is
possible to extend considerations to a more general case in which
$\Psi(\phi,\mathcal{B}) = \bar{C}^\al \chi_\al(\phi,\mathcal{B})$,
where the gauge-fixing functions $\chi_\al(\phi,\mathcal{B})$ depend
on \textit{all} the fields under consideration and satisfy the conditions
$\vp(\chi_\al)=\vp_{\al}$ and ${\rm gh}(\chi_\al)=0$. On the other
hand, the presence of the background field symmetry is not immediate
--- especially in the case of non-linear gauges --- as the gauge-fixing
functionals depend on the background fields. In what follows we
derive necessary conditions that the fermion gauge-fixing functional
should satisfy to achieve the consistent application of the BFM. As
it has been already mentioned above, during this process we shall
also find restrictions on the form of the generators $R^i_\al$ of
gauge symmetry.

Let us extend the transformation rule~\eqref{BGTrans} to the whole
set of quantum fields, as
\beq
\label{BGtrans_other}
\quad \de_\om^{(q)} B^\al
= -F^{\al}_{\ga\be}  B^\be \om^\ga ,
\quad
\de_\om^{(q)} C^\al = -F^{\al}_{\ga\be}  C^\be \om^\ga
 (-1)^{\vp_{\ga}} ,
 \quad \de_\om^{(q)} \bar{C}^\al
= -F^{\al}_{\ga\be}  \bar{C}^\be \om^\ga  (-1)^{\vp_{\ga}}.
\mbox{\,\,}
\eeq
Following the procedure used for the BRST symmetry, one can define the
operator of background field transformations,
\beq
\label{R_omega}
\hat{{R}}_\om (\phi,\mathcal{B}) \, = \,
\frac{\de_r}{\de \mathcal{B}^i}  \,\de^{(c)}_\om \mathcal{B}^i
\,+\,  \frac{\de_r}{\de \phi^A} \,
\de^{(q)}_\om \phi^A , \qquad \vp(\hat{{R}}_\om) \,=\, 0 .
\eeq
The gauge invariance of the initial classical action implies that
$\hat{{R}}_\om (\phi,\mathcal{B}) \, S_{0} (A+\mathcal{B}) = 0$.
Furthermore, it is not difficult to verify that the background gauge
operator commutes with the generator of BRST transformations,
\textit{i.e.},
\beq
[\hat{R}_{\text{BRST}} , \hat{{R}}_\om ] = 0.
\eeq
Combining this result with the representation~\eqref{FP-Psi} of the
Faddeev-Popov action, we get
\beq
\label{Condition}
\de_\om S_{\text{FP}}(\phi,\mathcal{B})
= \hat{{R}}_\om (\phi,\mathcal{B})
\, S_{\text{FP}}(\phi,\mathcal{B}) = 0
\quad
\Longleftrightarrow \quad \hat{{R}}_\om (\phi,\mathcal{B})
\, \Psi(\phi,\mathcal{B}) = 0 .
\eeq
In other words, the Faddeev-Popov action is invariant under background
field transformations if and only if the fermion gauge-fixing functional is
a scalar with respect to this transformation.

The condition~\eqref{Condition} constrains the possible forms of
the (extended) gauge-fixing function $\chi_\al(\phi,\mathcal{B})$,
as the relation
\beq
\hat{{R}}_\om (\phi,\mathcal{B}) \, \Psi(\phi,\mathcal{B}) =
\bar{C}^\al \de_\om \chi_\al(\phi,\mathcal{B}) -
F^{\al}_{\ga\be} \bar{C}^\be \om^\ga  (-1)^{\vp_{\ga}}
\chi_\al(\phi,\mathcal{B}) \,=\, 0
\eeq
fixes the transformation law for $\chi_\al(\phi,\mathcal{B})$,
\beq
\label{conditionChi}
\de_\om \chi_\al(\phi,\mathcal{B})
\,=\,  - \chi_\be (\phi,\mathcal{B})  F^{\be}_{\al\ga} \om^\ga.
\eeq
Therefore, in order to have the invariance of the Faddeev-Popov action
under  background field transformations it is necessary that the
gauge function $\chi_\al$ transforms as a tensor with respect to the
gauge group.
This requirement can be fulfilled provided that
$\chi_\al(\phi,\mathcal{B})$ is constructed only by using tensor
quantities.
Thus, Eq.~\eqref{conditionChi} may impose a restriction on the form
of gauge-fixing functions which are non-linear on the fields $A^i$.
Let us anticipate that an example of this kind will be presented in
Sec.~\ref{Sec.4}.

In order to complete the geometric point of view on the BFM
for gauge theories, let us introduce the generators
$\hat{\mathcal{R}}_\al = \hat{\mathcal{R}}_\al(\phi,\mathcal{B})$
of the background field transformation, defined by the rule
\beq
\hat{{R}}_\om(\phi,\mathcal{B})
\,=\, \hat{\mathcal{R}}_\al(\phi,\mathcal{B}) \om^\al ,
\qquad \vp(\hat{\mathcal{R}}_\al) = \vp_\al
\eeq
It is possible to verify that the generators of BRST and background
field transformation satisfy the relations
\beq
\hat{R}_{\text{BRST}}^2=0,
\qquad
[\hat{R}_{\text{BRST}} , \hat{\mathcal{R}}_\al ] = 0,
\qquad
[\hat{\mathcal{R}}_\al,\hat{\mathcal{R}}_\be]
= -  \hat{\mathcal{R}}_\ga F^\ga_{\be\al},
\label{alg}
\eeq
that define a superalgebra of the symmetries underlying the BFM.
 It is important to note that the last relation in (\ref{alg})
reproduces the gauge algebra~\eqref{AlgebraAi}, when restricted to the
sector of fields $\mathcal{B}^i$, and provides its generalization to the
whole set of fields $\phi^A$.

At this point we can conclude that
\beq
\hat{R}_{\text{BRST}}
(\phi,\mathcal{B}) \, S_{\text{FP}}(\phi,\mathcal{B}) = 0 \qquad
\text{and} \qquad \hat{\mathcal{R}}_\al
(\phi,\mathcal{B})\,\Psi(\phi,\mathcal{B}) = 0 \label{2conds}
\eeq
represent necessary conditions for the consistent application of the
BFM. The first relation is associated to the validity of the Ward
identity
(Slavnov-Taylor identities in the
case of Yang-Mills fields)
and the gauge independence of the vacuum functional
\footnote{
Let us mention that  the gauge independence of the vacuum functional
is needed for the gauge independent $S$-matrix and hence is a very
important element for the consistent quantum formulation of a gauge
theory \cite{BV1,KT}.},
while the second relation is called to provide the invariance of the
effective action in the BFM with respect to deformed (in the general
case) background field transformations. In what follows we shall
consider these statements explicitly. To this end, it is convenient
to introduce the extended action
\beq
S_\text{ext} (\phi,\mathcal{B},\phi^\ast)
\,=\,
S_\text{FP} (\phi,\mathcal{B})
+ \phi^\ast_A \, \hat{R}_{\text{BRST}} (\phi,\mathcal{B})
\, \phi^A ,
\eeq
where $\phi^\ast = \lbrace \phi^{\ast}_{ A}\rbrace$ denote the set
of sources (antifields) to the BRST transformations, with the parities
$\vp(\phi^{\ast}_{ A})=\vp_A + 1$. The corresponding (extended)
generating functional of Green functions reads
\beq
\label{Z_ext}
Z(\mathcal{J},\mathcal{B},\phi^\ast)
= \int \mathcal{D}\phi \exp\left\lbrace i
\left[  S_\text{FP} (\phi,\mathcal{B})  + \mathcal{J}_A \phi^A
+ \phi^\ast_A \, \hat{R}_{\text{BRST}} (\phi,\mathcal{B})
\, \phi^A  \right]  \right\rbrace ,
\eeq
where
$\mathcal{J}_A
= \big( J_i  , \,\, J^{(B)}_{\al}, \,\, \bar{J}_\al , \,\, J_\al \big)$
(with the parities $\vp(\mathcal{J}_A)=\vp_A$)
are the external sources for the fields $\phi^A$. The BRST symmetry,
together with the requirement that the generators $R^i_\al$ of gauge
transformation satisfy
\beq
\label{Jaco}
(-1)^{\vp_i} \frac{\de_\ell R^i_\al ( A+\mathcal{B})}{\de A^i}
+ (-1)^{\vp_\be + 1} F^\be_{\be\al}  = 0,
\eeq
implies the Ward identity
\beq
\label{Slavnov-Taylor}
\mathcal{J}_{A} \frac{\delta _{\ell}Z(
\mathcal{J},\mathcal{B},\phi ^{\ast })}{\delta \phi ^{\ast }_{ A}}
 \,=\,0.
\eeq
The relation~\eqref{Jaco} plays an important  role in the derivation
of the Ward identity insomuch as it ensures the triviality of the
Berezenian related to the change of integration variables in the form
of BRST transformations. In Yang-Mills theories, for instance, the
relations~\eqref{Jaco} are satisfied due to antisymmetry properties of
the structure constants.

The generating functional $W(\mathcal{J},\mathcal{B},\phi^\ast)$
of connected Green functions is defined in a usual way,
\beq
Z(\mathcal{J},\mathcal{B},\phi^\ast)
\,=\,e^{iW(\mathcal{J},\mathcal{B},\phi^\ast)},
\eeq
and the identity~\eqref{Slavnov-Taylor} can be cast into the form
\beq \label{Slavnov-TaylorW}
 \mathcal{J}_A \,\frac{\de_\ell  W(\mathcal{J},\mathcal{B},\phi^\ast)}{\de \phi^{\ast}_{ A}}
 \,=\,0.
\eeq
In order to construct the effective action (generating functional of
vertex functions), let us introduce the set of mean fields $\Phi = \{ \Phi^A \}$ with
\beq
\Phi^A
\,=\,
\left(  \mathfrak{A}^i, \mathfrak{B}^\al, \mathfrak{C}^\al ,
\bar{\mathfrak{C}}^\al \right),
\label{means}
\eeq
respectively for the fields
$A^i$, $B^\al$,  $C^\al$ and $\bar{C}^\al$, such that
\beq \label{means2}
\Phi^A
\,=\,
\frac{\de_\ell W(\mathcal{J},\mathcal{B},\phi^\ast)}{\de \mathcal{J}_A}.
\eeq
The (extended) effective action is defined as
\beq
\Ga \,=\,  \Gamma(\Phi,\phi^\ast,\mathcal{B}) \,=\,
W(\mathcal{J},\phi^\ast,\mathcal{B}) - \mathcal{J}_A \Phi^A
\eeq
and it satisfies the Ward identity
\beq
\left( \Gamma,\Gamma\right)  = 0 ,
\eeq
with the anti-bracket of two functionals defined by the rule
\cite{BV1,BV2}
\beq
(G,H) \,=\, \frac{\de_r G}{\de \phi^A}
\frac{\de_\ell H}{\de \phi^{\ast}_A}
- \frac{\de_\ell G}{\de \phi^{\ast}_A} \frac{\de_r H}{\de \phi^A}.
\eeq


Now, in order to explore the background field symmetry
it is useful to switch off the antifields and deal
with the traditional generating functions, \textit{e.g.},
\beq
Z(\mathcal{J},\mathcal{B}) \,=\,
Z(\mathcal{J},\mathcal{B},\phi^\ast)\vert_{\phi^\ast_A = 0}.
\label{Zz}
\eeq
Then, let us perform the
change of the functional integration variables
$\phi^A \mapsto \varphi^A (\phi)= \phi^A + \de_\om^{(q)} \phi^A$ in~\eqref{Z_ext}.
If the Berezenian associated to this change of variables is $\text{Ber}(\varphi) = 1$,
it follows
\beq \label{2108}
\int
\mathcal{D}\phi \left[ \de_\om S_{\text{FP}}(\phi,\mathcal{B})
-  \de^{(c)}_\om S_{\text{FP}}(\phi,\mathcal{B})
+ \mathcal{J}_A\de_\om^{(q)}\phi^A  \right]
\exp\left\lbrace i \left[  S_{\text{FP}}(\phi,\mathcal{B})
+ \mathcal{J}_A\phi^A \right]  \right\rbrace  \,=\, 0 .
\eeq

It is clear that if the generators $R^i_\al$ of gauge transformations
are linear, then $\text{Ber}(\varphi) = 1$ trivially when using the
dimensional regularisation, as the change of variables is linear.
On the other hand, in the more general scenario with
non-linear generators we meet a situation similar to that
of BRST transformations, where it was necessary to impose the
condition~\eqref{Jaco} to ensure the triviality of the Berezenian.
At the first sight it seems that the background field symmetry would
introduce further requirements on the generators of gauge
transformations. Nonetheless, when computing $\text{Ber}(\varphi)$ one finds out
that the contribution from the field $B^\al$ compensates the one
from $C^\al$ (or $\bar{C}^\al$, which transforms under the same rule), leading to
\beq
\text{Ber}(\varphi) = \exp \bigg[ \text{sTr} \bigg( \frac{\de_\ell \,
\de_\om^{(q)} \phi^A}{\de \phi^B} \bigg)  \bigg] = \exp \bigg[  (-1)^{\vp_i}
\frac{\de_\ell R^i_\al ( A+\mathcal{B})}{\de A^i} \om^\al
+ (-1)^{\vp_\be + 1} F^\be_{\be\al} \om^\al \bigg].
\eeq
Therefore, the condition~\eqref{Jaco} also ensures the triviality of the Berezenian
for the change of variables
$\phi^A \mapsto \phi^A + \de_\om^{(q)} \phi^A$.
This remarkable result means that, as it will be clear from
what follows, for the gauge theories considered
here the consistent use of the BFM does not impose an additional
requirement on the generators besides those that are already needed
in the framework of traditional approach (\textit{i.e.}, without using
the BFM).

After this important digression on the Berezenian, let us rewrite Eq.~\eqref{2108}
in the form
\beq
\label{ST-back-Z1}
&&
\de^{(c)}_\om Z(\mathcal{J},\mathcal{B})
\,=\,
i \mathcal{J}_A {T}^A_\om \left( \frac{\de_\ell}{\de (i\mathcal{J})},
\mathcal{B}\right)   Z(\mathcal{J},\mathcal{B})
\nonumber
\\
&+&
\int \mathcal{D}\phi \left[ i \hat{R}_{\text{BRST}}(\phi,\mathcal{B})
\hat{{R}}_\om(\phi,\mathcal{B})  \Psi(\phi,\mathcal{B})\right]
\exp\left\lbrace i \left[  S_{\text{FP}}(\phi,\mathcal{B})
+ \mathcal{J}_A\phi^A \right]  \right\rbrace ,
\eeq
where we introduced the notation
\beq
T^A_\om(\phi,\mathcal{B})
\,=\, \left( \left[  R^j_\al ( A +\mathcal{B})
- R^j_\al (\mathcal{B}) \right]\om^\al   , \,
F^\al_{\be\ga} \om^\ga B^\be  ,  \,
F^\al_{\be\ga} \om^\ga C^\be , \,
F^\al_{\be\ga} \om^\ga \bar{C}^\be \right) .
\eeq
If the condition
$ \hat{{R}}_\om(\phi,\mathcal{B})  \Psi(\phi,\mathcal{B})=0$
(see Eq.~\eqref{Condition}) for the invariance of the Faddeev-Popov
action with respect to the background field transformations is
satisfied, Eq.~\eqref{ST-back-Z1} can be written in the closed form
\beq
\label{ST-back-Z2}
\de^{(c)}_\om Z(\mathcal{J},\mathcal{B})
\, = \, i \mathcal{J}_A {T}^A_\om
\Big( \frac{\de_\ell}{\de (i\mathcal{J})},\mathcal{B}\Big)
Z(\mathcal{J},\mathcal{B}),
\eeq
or, in terms of the  generating functional of connected Green functions,
\beq \label{ST-back-W}
\de^{(c)}_\om W(\mathcal{J},\mathcal{B})
 \,=\,  \mathcal{J}_A {T}^A_\om \left(
 \frac{\de_\ell W(\mathcal{J},\mathcal{B})}{\de \mathcal{J}}
 + \frac{\de_\ell }{\de (i\mathcal{J})},\mathcal{B}\right) \cdot 1.
\eeq
Here $1$ is the unite vector in the space of the fields.

Finally, for the effective action, the invariance of the Faddeev-Popov
action with respect to background field transformations implies that
\beq
\label{GammaR}
\hat{{R}}_\om (\Phi,\mathcal{B}) \, \Gamma(\Phi,\mathcal{B})= 0 ,
\eeq
where $\hat{{R}}_\om (\Phi,\mathcal{B})$ is defined by the
substitution $\phi^A \mapsto \Phi^A$ in Eq.~\eqref{R_omega}
with the (deformed) transformations of the mean fields
(\textit{cf.}~\eqref{BGTrans} and~\eqref{BGtrans_other})
\beq
\label{TransMeanF}
\de_\om^{(q)} \mathfrak{A}^i
\,=\, \langle R^i_\al \rangle (\mathfrak{A}, \mathcal{B})  \om^\al ,
\quad\,
&&
\quad \de_\om^{(q)} \mathfrak{B}^{\al}
\,=\, - F^\al_{\ga\be} \mathfrak{B}^\be \om^\ga ,
\\
\nonumber
\quad \de_\om^{(q)} \mathfrak{C}^\al
= -F^\al_{\ga\be} \mathfrak{C}^\be \om^\ga (-1)^{\vp_{\ga}},
&&\quad \de_\om^{(q)} \bar{\mathfrak{C}}^\al
= -F^\al_{\ga\be} \bar{\mathfrak{C}}^\be \om^\ga (-1)^{\vp_{\ga}} .
\eeq
In the sector of fields $\mathfrak{A}^i$, the generator is given by
\beq
\langle R^i_\al \rangle (\mathfrak{A}, \mathcal{B})
\,=\,  R^i_\al (\hat{\mathfrak{A}} + \mathcal{B})
\cdot 1 - R^i_\al(\mathcal{B}),
\qquad
\text{with}
\qquad
\hat{\Phi}^A \,=\,
\Phi^A + i \left( \Ga^{\prime\prime -1}\right)^{AB}
\frac{\de_\ell }{\de \Phi^B},
\eeq
where $\Ga^{\prime\prime -1}$ is the inverse matrix of second
derivatives of the effective action,  
\beq
\big(
\Ga^{\prime\prime -1}\big)^{AC} \Ga^{\prime\prime}_{CB}
= \de^A_B,
\qquad
\mbox{where}
\qquad
\Ga^{\prime\prime}_{AB}(\Phi,\mathcal{B})
\,=\, \frac{\de_\ell}{\de \Phi^A}
\frac{\de_r \Ga(\Phi,\mathcal{B})}{\de \Phi^B}.
\eeq
Let us stress that the relation \eqref{GammaR} follows
from~\eqref{ST-back-Z2}, which only holds if the condition
$\hat{{R}}_\om(\phi,\mathcal{B}) \, \Psi(\phi,\mathcal{B})=0$ is
satisfied. The property~\eqref{GammaR} is crucial to the BFM
inasmuch as, when the mean fields $\Phi$ are switched off, it ensures
the gauge invariance of the  functional
$\Ga(\mathcal{B})=\Ga(\Phi,\mathcal{B})\vert_{\Phi^A = 0}$, namely,
\beq
\de_\om^{(c)} \Ga(\mathcal{B}) = 0.
\eeq

It is worth mentioning that the previous expressions can be simplified
if the generators $R^i_\al$ of gauge transformations are linear in fields
$A^i$. In this case one can define the background field transformation
of  all the external sources as
\beq
\de_\om^{(q)} J_i
\,=\, - J_j R^j_{\al,i} (A) \om^\al (-1)^{\vp_\al \vp_i} ,
&&
\quad \de_\om^{(q)} J^{(B)}_{\al} = - J^{(B)}_{\be} F^\be_{\al\ga} \om^\ga ,
\\
\nonumber
\quad \de_\om^{(q)} \bar{J}_\al =- \bar{J}_\be F^\be_{\al\ga} \om^\ga  ,
&&\quad \de_\om^{(q)} J_\al = - J_\be F^\be_{\al\ga} \om^\ga .
\eeq
Then, Eqs.~\eqref{ST-back-Z2} and~\eqref{ST-back-W}  boil down to
\beq
\label{ST-back-Z3}
\de^{(c)}_\om Z(\mathcal{J},\mathcal{B})
+ \de^{(q)}_\om \mathcal{J}_A
\frac{\de_\ell Z(\mathcal{J},\mathcal{B}) }{\de \mathcal{J}_A}\,=\, 0
\eeq
and
\beq
\label{ST-back-W3}
\de^{(c)}_\om W(\mathcal{J},\mathcal{B})
+ \de^{(q)}_\om \mathcal{J}_A
\frac{\de_\ell W(\mathcal{J},\mathcal{B}) }{\de \mathcal{J}_A}
\,=\, 0.
\eeq
In other words, the invariance of the Faddeev-Popov action with
respect to the background field transformations implies the
invariance of the generating functionals of Green functions with
respect to the joint transformation $\de_\om^{(q)}\mathcal{J}_i$
of the external sources and $\de_\om^{(c)}\mathcal{B}$ of the
background field. Moreover, in what concerns the effective action,
under this circumstance the transformation~\eqref{TransMeanF}
of the mean fields $\mathfrak{A}^i$ are not deformed, \textit{i.e.},
\beq
\de_\om^{(q)} \mathfrak{A}^i \,=\, [R^i_\al(\mathfrak{A}
+ \mathcal{B}) - R^i_\al(\mathcal{B})]  \om^\al.
\eeq

For the sake of completeness, let us compare the generating
functionals  in the background field formalism and in the traditional
one --- and, ultimately, their relations with $\Ga(\mathcal{B})$.
Consider the generating functional of Green functions which
corresponds to the standard quantum field theory approach, but in a
 very special gauge fixing,
\beq
\label{Z2}
Z_2(\mathcal{J})
\,=\,  \int \mathcal{D}\phi \exp\left\lbrace
i \left[  S_{0} (A) + \hat{R}_{\text{BRST}}(\phi) \,
\Psi(A-\mathcal{B},B,C,\bar{C},\mathcal{B})
+ \mathcal{J}_A \phi^A \right]  \right\rbrace .
\eeq
In the last expression all the dependence of the quantity
$Z_2(\mathcal{J})$ on the external field is only through the
gauge-fixing functional. Thus, this functional depends the external
field $\mathcal{B}^i$, but since this dependence is not of the
BFM type, $Z_2(\mathcal{J})$ is nothing else but the
conventional generating functional of Green functions of the theory,
defined by $S_0$ in a specific $\mathcal{B}^i$-dependent gauge.
One of the consequences is that any kind of physical results does
not depend on  $\mathcal{B}^i$. Furthermore, in Eq.~(\ref{Z2})
$\hat{R}_{\text{BRST}}(\phi)$ is defined by setting
$\mathcal{B}^i = 0$ in Eq.~\eqref{R^i_B}. The arguments of
$\Psi$ are written explicitly, showing that we assume that $A^i$
only occurs in a specific combination with $\mathcal{B}^i$.
We stress that, being formulated in the traditional way (\textit{i.e.}, not
in the BFM), $\,Z_2(\mathcal{J})\,$ \textit{per se} does not impose
any constraint on the linearity of the gauge-fixing fermion $\Psi$
with respect to the quantum field $A^i$.

Making some change of variables in the functional integral, it is
easy to verify that (\ref{Zz}) is related to (\ref{Z2}) in the
following way:
\beq
Z(\mathcal{J},\mathcal{B})
\,=\, Z_2(\mathcal{J}) \exp \left( - i J_i \mathcal{B}^i \right) .
\eeq
Accordingly, for the generating functional of connected Green
functions one has
\beq
\label{W}
W(\mathcal{J},\mathcal{B}) = W_2(\mathcal{J}) - J_i \mathcal{B}^i,
\eeq
where $Z_2(\mathcal{J}) = e^{i W_2(\mathcal{J})}$.
Recall that, according to (\ref{means2}),
\beq
\label{MeanField}
\mathfrak{A}^i  \,=\, \frac{\de_\ell W(\mathcal{J},\mathcal{B})}{\de J_i}.
\eeq
Similarly,
\beq
\mathfrak{A}_{2}^i \equiv
\frac{\delta _{\ell } W_2(\mathcal{J})}{\de J_i}
\,=\, \mathfrak{A}^i + \mathcal{B}^i
\eeq
Following the same line, let us define the effective action associated
to $Z_2(\mathcal{J})$, as
\beq
\Ga_2(\mathfrak{A}_2^i,\mathfrak{B}^\al,
\mathfrak{C}^\al , \bar{\mathfrak{C}}^\al )
\,=\, W_2(\mathcal{J}) - J_i\mathfrak{A}_{2}^i
- J ^{(B)}_\al\mathfrak{B}^\al - \bar{J}_\al \mathfrak{C}^\al
- J_\al \bar{\mathfrak{C}}^\al.
\eeq
A moment's reflection shows that
\beq
\label{55}
\Ga(\Phi,\mathcal{B})
\,=\, \Ga_2(\mathfrak{A}_2^i,\mathfrak{B}^\al,
\mathfrak{C}^\al , \bar{\mathfrak{C}}^\al).
\eeq
In other words, the effective action $\Ga(\Phi,\mathcal{B})$ in the
background field formalism is equal to the initial effective action
in a particular gauge with mean field
$\,\mathfrak{A}_2^i = \mathfrak{A}^i + \mathcal{B}^i$ --- or,
switching off the mean fields, $\Ga(\mathcal{B})
= \Ga_2(\mathfrak{A}_2)\vert_{\mathfrak{A}_2^i=\mathcal{B}^i}$.
We point out that the aforementioned particularity of the gauge is
not associated to its linearity with respect to the quantum fields,
but to its dependence on $\mathcal{B}$ (see Eq.~\eqref{Z2}).

\section{Two particular cases}
\label{Sec.4}

In this section we present the applications of the formalism
described above, to the Yang-Mills and quantum gravity theories.

\subsection{Yang-Mills theory}

As an example of the results presented in the previous section, let
us consider the case of the pure Yang-Mills theory, defined by the
action
\beq
\label{ActionYM}
S_{0} (A) = - \frac{1}{4} F^a_{\mu\nu}(A) F^a_{\mu\nu} (A),
\eeq
where
$\,F^a_{\mu\nu}(A) = \pa_\mu A^a_\nu - \pa_\nu A^a_\mu
+ f^{abc} A^b_\mu A^c_\nu\,$
is the field strength for the non-Abelian  vector field $A_\mu$,
taking values in the adjoint representation of  a compact semi-simple
Lie group. Being a particular case of the more general theory
described above, it is instructive to present the correspondence
with the notations used in Sec.~\ref{Sec.3}, namely
\beq
A^i \mapsto A^a_\mu ,
\qquad
\mathcal{B}^i \mapsto \mathcal{B}^a_\mu ,
\qquad
F^{\al}_{\be\ga} \mapsto f^{abc} ,
\qquad R^i_\al(A) \mapsto D^{a b}_\mu(A)
= \de^{ab} \partial_\mu + f^{acb} A^c_\mu.
\eeq
Here the structure coefficients $f^{abc}$ of the gauge group are
constant. The action~\eqref{ActionYM} is invariant under the gauge
transformations defined by the generator $D^{ab}_\mu (A)$ with an
arbitrary gauge function $\om^a$ with $\vp(\om^a) =0$. In the
Faddeev-Popov quantization, the Grassmann parity of the fields
$\,\phi^A = ( A^a_\mu, B^a, C^a, \bar{C}^a )$ is, respectively,
$\vp_A = (0,0,1,1)$.

The background field formalism for Yang-Mills theory comprises
the definition of the background field transformation (see
Eqs.~\eqref{BGTrans} and~\eqref{BGtrans_other})
\beq \label{YM-back}
&& \de_\om^{(q)} A^a_\mu \,=\, f^{abc} A^b_\mu \om^c ,
\qquad
\de_\om^{(c)} \mathcal{B}^a_\mu = D^{ab}_\mu (\mathcal{B}) \om^b ,
\nonumber
\\
&&
\de_\om^{(q)} B^a = f^{abc}  B^b \om^c ,
\qquad
\de_\om^{(q)} C^a = f^{abc} C^b \om^c ,
\qquad
\de_\om^{(q)} \bar{C}^a = f^{abc} \bar{C}^b \om^c  .
\mbox{\qquad}
\eeq
Note that, in agreement to~\eqref{BGTrans},  the generator of the
 transformation in the sector of fields $A^a_\mu$ reads
\beq
D^{ab}_\mu (A + \mathcal{B}) - D^{ab}_\mu (\mathcal{B})
= f^{acb} A^c ,
\eeq
and thus all the quantum fields transform according the same rule.
Also, the condition~\eqref{Condition} for the background field
invariance of the Faddeev-Popov action reads
\beq \label{conditionChiYM}
\hat{{R}}_\om (\phi,\mathcal{B}) \chi^a(A,B,\mathcal{B})
\,=\, f^{abc} \chi^b (A,B,\mathcal{B}) \om^c.
\eeq
This requirement can be fulfilled provided that
$\,\chi^a(A,B,\mathcal{B})\,$ is constructed only by using vectors
(in the group index) such as $B^a$, $A^a_\mu$,
$F^a_{\mu\nu}(\mathcal{B})$, $D^{ab}_\mu (\mathcal{B}) A^b_\nu$,
$D^{ab}_\mu(\mathcal{B}) D^{bc}_\nu (\mathcal{B}) A^c_\al$ and so
on. The most simple gauge-fixing function compatible  with this
condition is
\beq
\chi^a (A,B,\mathcal{B}) = D^{ab}_\mu (\mathcal{B}) A^b_{\mu},
\eeq
that turns out to be the most popular gauge-fixing function for Yang-Mills
theory. In principle, nonetheless, non-linear gauges which
satisfy~\eqref{conditionChiYM} can also be used. One simple example
is
\beq
\label{ExampleGauge}
\chi^a(A,B,\mathcal{B}) \,=\,
F^a_{\mu\nu}(\mathcal{B}) \left[ D^{bc}_\mu (\mathcal{B})
A^c_\nu \right] \left[ D^{bd}_\lambda (\mathcal{B}) A^d_\lambda
\right],
\eeq
which is quadratic on the quantum field. Note, however, that the
gauge resultant from the substitution of $F^a_{\mu\nu}(\mathcal{B})$
by $F^a_{\mu\nu}(A)$ in~\eqref{ExampleGauge} is not admissible,
as it violates the transformation law~\eqref{conditionChiYM}. We
also point out that  the specific dependence of the gauge condition
on the auxiliary field $B^a$ (and also on the ghost fields, in the
generalisation proposed in Sec.~\ref{Sec.3}) is not critical, as it
already satisfies the vector transformation law.


We conclude that, in principle, it is possible to use
non-linear gauge fixing conditions for the quantum fields in
Yang-Mills theories formulated in the BFM, provided that the
relation~\eqref{conditionChiYM} is satisfied. Such procedure may
introduce parameters with negative mass dimension, which could
affect renormalization; however, a general analysis on this issue
is beyond the scope of the present work.

\subsection{Quantum gravity}

As a second example, consider the case of quantum gravity theories, defined by the
action $S_{0} (g)$ of a Riemann metric $g = \lbrace g_{\mu\nu}(x)\rbrace$ with $\vp(g)=0$,
and which is invariant under general coordinate transformations.
The generator of such transformation is linear and reads
\beq
R_{\mu\nu\sigma}(x,y;g) = - \delta(x-y) \partial_\sigma g_{\mu\nu}(x) -
g_{\mu\sigma}(x) \partial_\nu \delta(x-y) - g_{\sigma\nu}(x) \partial_\mu \delta(x-y).
\eeq
Therefore, for an arbitrary gauge function $\om^\alpha$ with $\vp(\om^\alpha) =0$
one has $\delta_\omega g_{\mu\nu} = R_{\mu\nu\sigma}(g) \omega^\sigma$, or,
writing all the arguments explicitly,
\beq
\delta_\omega g_{\mu\nu}(x) = \int dy \, R_{\mu\nu\sigma}(x,y;g) \omega^\sigma(y).
\eeq
In this case, the structure functions are given by
\beq
F^{\al}_{\be\ga}(x,y,z) = \delta(x-y)\delta^\alpha_\gamma \partial^{(x)}_\be \delta(x-z)
 - \delta(x-z) \delta^\al_\be \partial^{(x)}_\ga \delta(x-y),
\eeq
which satisfy $F^{\al}_{\be\ga}(x,y,z) = - F^{\al}_{\ga\be}(x,z,y)$, as usual.

In terms of the notation used in Sec.~\ref{Sec.3} one has the correspondence
\beq
A^i \mapsto g_{\mu\nu}(x) ,
\qquad
\mathcal{B}^i \mapsto \bar{g}_{\mu\nu}(x) ,
\qquad R^i_\al(A) \mapsto R_{\mu\nu\sigma}(x,y;g) ,
\qquad
F^{\al}_{\be\ga} \mapsto F^{\al}_{\be\ga}(x,y,z).
\eeq
According to~\eqref{BGTrans} and~\eqref{BGtrans_other}, in the BFM one defines
 the background field transformation
\beq
\de_\om^{(c)} \bar{g}_{\mu\nu} &=&   R_{\mu\nu\sigma}(\bar{g}) \om^\sigma \, =
\, -\om^\si \partial_\si \bar{g}_{\mu\nu} - \bar{g}_{\mu\si} \partial_\nu \om^\si -
\bar{g}_{\si\nu} \partial_\mu \om^\si ,
\\
\de_\om^{(q)} g_{\mu\nu} &=&  [ R_{\mu\nu\sigma}(g+\bar{g}) -
R_{\mu\nu\sigma}(\bar{g})] \om^\sigma \, = \, R_{\mu\nu\sigma}(g) \om^\si ,
\\
\de_\om^{(q)} V^\al &= & - F^{\al}_{\be\ga}  V^\ga \om^\be
\,=\, -\omega^\si \partial_\si V^\al + V^\si \partial_\si \omega^\al,
\eeq
where $V^\al$ denotes any of the vector fields $C^\al$, $\bar{C}^\al$ and $B^\al$.
Thence, the condition~\eqref{Condition} for the background field invariance
of the Faddeev-Popov action now reads
$\delta_\om \chi_\alpha =
- \om^\si \partial_\si \chi_\al - \chi_\si \partial_\al \om^\si -
\chi_\al \partial_\si \om^\si$, \textit{i.e.}, the gauge-fixing function
must transform as a vector density. Writing the density part explicitly,
as it is standard for gravity theories, in the BFM one has $\chi_\al(\phi,\bar{g})=
\sqrt{-\det\bar{g}} \, \tilde{\chi}_\al (\phi,\bar{g})$, with $\tilde{\chi}_\al$
transforming as a vector field,
\beq \label{qgc}
\delta_\om \tilde{\chi}_\al = - \om^\si \partial_\si \tilde{\chi}_\al -
\tilde{\chi}_\si \partial_\al \om^\si.
\eeq

Differently from the case of Yang-Mills theory (see~\eqref{YM-back}),
in gravity theories all the fields under consideration transform as tensor ones
with respect to the group algebra. Therefore, the condition~\eqref{qgc}
is automatically satisfied for any gauge-fixing condition, including
non-linear ones.
Furthermore, since in gravity the coupling constant has negative dimension,
it is natural that non-linear gauge and field parametrizations have fruitful
applications in these models (see, \textit{e.g.},~\cite{Ven}). Indeed,
the gauge that is linear in one parametrization becomes non-linear in
another one, as one can observe, \textit{e.g.}, in the recent paper \cite{J-QG}.

\section{Conclusions}
\label{Sec.5}

We introduced a new framework for formulating the BFM in a general
type of Yang-Mills theories. The new approach is based on identifying the
superalgebra (\ref{alg}) composed by the generators of BRST and
background field transformations. It is shown that the condition for
consistent use of the BFM is that the gauge-fixing fermion functional
$\Psi$ must be a scalar with respect to the background field
transformation. This condition produces a restriction on the
admissible forms of the non-linear gauge-fixing condition used
within the BFM. At the same time, these restrictions open the way
for using a wide class of gauge-fixing conditions, including
the gauges which are non-linear with respect to quantum fields. The
considerations presented above can be directly extended to the case
when the gauge-fixing function $\chi_\al$ depends on all quantum
fields under consideration, including the ghost fields, as discussed
in Sec.~\ref{Sec.3}. As an application of general formulation, we
discussed the non-linear gauge conditions for the Yang-Mills
and  quantum gravity theories in the background field formalism.

Realistic models of fundamental interactions \cite{Weinberg} use the
mechanism of spontaneous symmetry breaking to generate the masses
of the physical particles \cite{Higgs}. Up to now there is no consistent
formulation of gauge theories with spontaneous symmetry breaking
within the BFM. All quantum studies of a such kind of gauge theories
are restricted to the one-loop approximation (see, \textit{e.g.},
early papers \cite{Niel,Shore} and recent ones \cite{DGM,CM}).
For example, an attempt to consider the spontaneous symmetry breaking
in the presence of external gravity may lead to serious complications
\cite{TmnBalt}. It is
clear that the consistent analysis of invariant renormalization in this
case is a very challenging problem.

\section*{Acknowledgements}
\noindent
B.L.G. and P.M.L. are grateful to the Department of Physics
of the Federal University of Juiz de Fora (MG, Brazil) for warm
hospitality during their long-term visits.  The work of P.M.L. is
supported partially by the Ministry of Science and Higher Education of the Russian Federation, grant  3.1386.2017 and by the RFBR grant
18-02-00153.
This work of I.L.Sh. was partially supported by Conselho Nacional de
Desenvolvimento Cient\'{i}fico e Tecnol\'{o}gico - CNPq under the
grant 303635/2018-5 and Funda\c{c}\~{a}o de Amparo \`a Pesquisa
de Minas Gerais - FAPEMIG under the project APQ-01205-16.



\begin{thebibliography}{99}

\bibitem{FP} L.D. Faddeev and V.N. Popov,
{\it Feynman diagrams for the Yang-Mills field},
Phys. Lett. {\bf B25} (1967) 29.

\bibitem{BV1} I.A. Batalin and G.A. Vilkovisky.
{\it Gauge algebra and quantization},
Phys. Lett. {\bf B102} (1981) 27.

\bibitem{BV2} I.A. Batalin and  G.A. Vilkovisky,
{\it Quantization of gauge theories with linearly dependent
generators},
Phys. Rev. {\bf D28} (1983) 2567.

\bibitem{RenCurved} P.M. Lavrov and I.L. Shapiro,
{\it On the renormalization of gauge theories in curved space-time,}
Phys. Rev.  {\bf D81} (2010) 044026,
arXiv: 0911.4579.

\bibitem{DeW} B.S. De Witt,
{\it Quantum theory of gravity. II. The manifestly covariant theory},
Phys. Rev. {\bf 162} (1967) 1195.

\bibitem{AFS}
I.Ya. Arefeva, L.D. Faddeev and A.A. Slavnov,
{\it Generating functional for the s matrix in gauge theories},
Theor. Math. Phys. {\bf 21} (1975) 1165
(Teor. Mat. Fiz. {\bf 21} (1974) 311-321).

\bibitem{Abbott}
L.F. Abbott, {\it The background field method beyond one loop},
Nucl. Phys.  {\bf B185} (1981) 189.

\bibitem{CountGhost} I.L. Shapiro,
{\it Counting ghosts in the ``ghost-free'' non-local gravity},
Phys. Lett.  {\bf B744} (2015) 67,
arXiv:1502.00106.

\bibitem{LSh}  P.M. Lavrov and I.L. Shapiro,
{\it Gauge invariant renormalizability of quantum gravity},
Phys.\ Rev.\ D {\bf 100} (2019)  026018,
arXiv:1902.04687.

\bibitem{K-SZ} H. Kluberg-Stern and J.B. Zuber,
{\it Renormalization of non-Abelian
gauge theories in a background-field gauge. I. Green's functions},
Phys. Rev. {\bf D12} (1975) 482.

\bibitem{GvanNW}
M.T. Grisaru, P. van Nieuwenhuizen and C.C. Wu,
{\it Background field method versus normal field theory in
explicit examples: One loop divergences in $S$-matrix and
Green's functions for Yang-Mills and gravitational fields},
Phys. Rev. {\bf D12} (1975) 3203.

\bibitem{CMacL} D.M. Capper and A. MacLean,
{\it The background field method at two loops:
A general gauge Yang-Mills calculation},
Nucl. Phys. {\bf B203} (1982) 413.

\bibitem{IO} S. Ichinose and M. Omote,
{\it Renormalization using the background-field formalism},
Nucl. Phys. {\bf B203} (1982) 221.

\bibitem{GS} M.H. Goroff and A. Sagnotti,
{\it The ultraviolet behavior of Einstein gravity},
Nucl. Phys. {\bf B266} (1986) 709.

\bibitem{Ven} A.E.M. van de Ven,
{\it Two-loop quantum gravity},
Nucl. Phys. {\bf B378} (1992) 309.

\bibitem{Reuter} M.~Reuter and C.~Wetterich,
{\it Effective average action for gauge theories and exact evolution
equations},
Nucl. Phys. {\bf B417} (1994) 181.

\bibitem{Gr} P.A. Grassi,
{\it Algebraic renormalization of Yang-Mills
theory with background field method},
Nucl. Phys. {\bf B462} (1996) 524.

\bibitem{BC} C. Becchi  and R. Collina,
{\it Further comments on the background field method and
gauge invariant effective action},
Nucl. Phys. {\bf B562} (1999) 412.

\bibitem{FPQ} R. Ferrari, M. Picariello and A. Quadri,
{\it Algebraic aspects of the background field method},
Annals Phys. {\bf 294} (2001) 165.

\bibitem{BQ} D. Binosi, A. Quadri,
{\it The background field method as a canonical transformation},
Phys. Rev. {\bf D85} (2012) 121702.

\bibitem{Barvinsky:2017zlx} A.O.~Barvinsky, D.~Blas,
M.~Herrero-Valea, S.M.~Sibiryakov and C.F.~Steinwachs,
{\it Renormalization of gauge theories in the background-field
approach},
JHEP {\bf 1807} (2018) 035,
arXiv:1705.03480.

 \bibitem{BLT-YM} I.A. Batalin, P.M. Lavrov and I.V. Tyutin,
{\it Multiplicative renormalization of Yang-Mills theories in the
background-field formalism},
Eur. Phys. J. {\bf C78} (2018) 570.

\bibitem{FT} J. Frenkel and J.C. Taylor,
{\it Background gauge renormalization and BRST identities},
Annals Phys. {\bf 389} (2018) 234.

\bibitem{Lav} P.M. Lavrov,
{\it Gauge (in)dependence and background field formalism},
Phys. Lett. {\bf B791} (2019) 293.

\bibitem{BFMc} F.T. Brandt,  J. Frenkel and D.G.C. McKeon,
{\it Renormalization of six-dimensional Yang-Mills theory in a
background gauge field},
Phys. Rev. {\bf D99} (2019) 025003.

\bibitem{BLT-YM2} I.A. Batalin, P.M. Lavrov and I.V. Tyutin,
{\it Gauge dependence and multiplicative renormalization of
Yang-Mills theory with matter fields},
Eur. Phys. J. {\bf C79} (2019) 628,
arXiv:1902.09532.

\bibitem{BRS1} C. Becchi, A. Rouet and R. Stora,
{\it The abelian Higgs Kibble Model, unitarity of the $S$-operator},
Phys. Lett. {\bf B52} (1974) 344.

\bibitem{T} I.V. Tyutin,
{\it Gauge invariance in field theory and statistical
physics in operator formalism}, Lebedev Inst. preprint
N 39 (1975).

\bibitem{Das:1980zy}   A.K.~Das,
{\it Nontrivial Quadratic Gauge Fixing in Yang-Mills Theories,}
Pramana {\bf 16} (1981) 409.

\bibitem{Gavela:1981ri}
  M.B.~Gavela, G.~Girardi, C.~Malleville and P.~Sorba,
{\it A Nonlinear R($\xi$) Gauge Condition for the Electroweak
$SU(2) \times U(1)$ Model,}
Nucl. Phys. {\bf B193} (1981) 257.

\bibitem{Gervais:1972tr}  J.L.~Gervais and A.~Neveu,
{\it Feynman rules for massive gauge fields with dual diagram
topology,}
Nucl. Phys. {\bf B46} (1972) 381.

\bibitem{Fujikawa:1973qs}
K.~Fujikawa,
{\it $\xi$-limiting process in spontaneously broken gauge theories,}
Phys. Rev. {\bf D7} (1973) 393.

\bibitem{Raval:2016nsz} H.~Raval,
{\it Absence of the Gribov ambiguity in a quadratic gauge,}
Eur. Phys. J. {\bf C76} (2016) 243,
arXiv:1604.00674.

\bibitem{Weinberg80} S.~Weinberg,
{\it Effective Gauge Theories,}
Phys. Lett.  {\bf B91} (1980) 51.

\bibitem{Hsu:1973yf} J.P.~Hsu,
{\it Unitarity in gauge theories with nonlinear gauge conditions,}
Phys. Rev.{\bf D8} (1973) 2609.

\bibitem{Shizuya:1976rz}   K.-i.~Shizuya,
{\it Renormalization of Gauge Theories with Nonlinear Gauge
Conditions,}
Nucl. Phys. {\bf B109} (1976) 397.

\bibitem{Tyutin:1981ws}  I.V.~Tyutin,
{\it Renormalization Of Gauge Theories In Nonlinear Gauges,}
Sov. Phys. J.  {\bf 24} (1981) 487.

\bibitem{Girardi:1982by} G.~Girardi, C.~Malleville and P.~Sorba,
{\it General Treatment of the Nonlinear R ($\epsilon$) Gauge
Condition,}
Phys. Lett.  {\bf B117} (1982) 64.
  
\bibitem{VLT82} B.L. Voronov, P.M. Lavrov, I.V. Tyutin,
{\it Canonical Transformations And The Gauge Dependence In General
Gauge Theories},
Sov. J. Nucl. Phys. {\bf 36} (1982) 498.

\bibitem{ZinnJustin:1984dt}  J.~Zinn-Justin,
{\it Renormalization of Gauge Theories: Nonlinear Gauges,}
Nucl. Phys.{\bf B246} (1984) 246.

\bibitem{GS85}   M.H.~Goroff and A.~Sagnotti,
 {\it Quantum Gravity At Two Loops,}
  Phys. Lett.  {\bf B160} (1985) 81.

\bibitem{Grassi:2003mv}   P.A.~Grassi, T.~Hurth and A.~Quadri,
{\it Super background field method for N=2 SYM,}
JHEP {\bf 0307} (2003) 008,
arXiv:hep-th/0305220.

\bibitem{Bornsen:2002hh} J.P.~Bornsen and A.E.M.~van de Ven,
{\it Three loop Yang-Mills beta function via the covariant background
field method,}
Nucl. Phys. {\bf B657} (2003) 257,
arXiv:hep-th/0211246.

\bibitem{DeWitt} B.S. DeWitt,
{\it Dynamical theory of groups and fields} (Gordon and Breach, 1965).

\bibitem{KT}
R.E. Kallosh and I.V. Tyutin,
{\it The equivalence theorem and gauge invariance in
renormalizable theories}, Sov. J. Nucl. Phys. {\bf 17} (1973) 98.

\bibitem{J-QG} J.D. Gon\c{c}alves, T. de Paula Netto and I.L. Shapiro,
{\it Gauge and parametrization ambiguity in quantum gravity,}
Phys. Rev. {\bf D97} (2018) 026015,
arXiv:1712.03338.

\bibitem{Weinberg} S. Weinberg,
{\it The Quantum theory of fields}, Vol.II
(Cambridge University Press, 1996).

\bibitem{Higgs}
P.W. Higgs, {\it Broken symmetries, massless particles and gauge fields},
Phys. Lett. {\bf 12} (1964) 132.

\bibitem{Niel}
N.K. Nielsen, {\it On the gauge dependence of spontaneous
symmetry breaking in gauge theories},
Nucl. Phys.  {\bf B101} (1975) 173.

\bibitem{Shore}
G.M. Shore, {\it Symmetry restoration and the background field method in gauge
theories}, Ann. Phys. {\bf 137} (1981) 261.

\bibitem{DGM}
M.-L. Du, F.-K. Guo and U.-G. Mei{\ss}ner,
{\it One-loop renormalization of the chiral Lagrangian
for spinless matter fields in the SU(N)
fundamental representation}, J. Phys. {\bf G44} (2017) 014001,
arXiv:1607.00822.

\bibitem{CM} O. Cat\`a and  C. M\"uller, {\it Chiral effective theories with
a light scalar at one loop},
arXiv:1906.01879.

\bibitem{TmnBalt}
M. Asorey, P.M. Lavrov, B.J. Ribeiro and I.L. Shapiro,
{\it Vacuum stress-tensor in SSB theories,}
Phys. Rev. {\bf D85} (2012) 104001,
arXiv:1202.4235.

\end{thebibliography}
\end{document}